\documentclass[aps,prd,reprint]{revtex4-2}

\usepackage{amsmath,amssymb,physics}
\usepackage{graphicx}
\usepackage{caption}
\usepackage{verbatim}

\usepackage[colorlinks=true,
            linkcolor=blue,
            citecolor=blue,
            urlcolor=blue]{hyperref}

\begin{document}

\title{Einstein–de Haas effect and induced rotation in an evolving magnetized QCD matter}

\author{Dushmanta Sahu}
\email{Dushmanta.Sahu@cern.ch}
\affiliation{Instituto de Ciencias Nucleares, UNAM, Apartado Postal 70-543, Coyoacán, 04510, México City, México}

\author{Captain R. Singh}
\email{captainriturajsingh@gmail.com}
\affiliation{Department of Physics, Indian Institute of Technology Indore, Simrol, Indore 453552, India}

\begin{abstract}
The Einstein-de Haas (EdH) effect describes the emergence of collective rotation driven by spin alignment under an external magnetic field. We investigate this effect in a dynamically expanding quark-gluon plasma (QGP) using a quasiparticle model (QPM). We compute the EdH-induced angular velocity $\omega_{\mathrm{EdH}}$ as a function of temperature, proper time, and fireball radius. Our results show that $\omega_{\mathrm{EdH}}$ grows with proper time and is consequently suppressed at higher temperatures. Near the QGP crossover temperature, $\omega_{\mathrm{EdH}}$ attains a substantial, non-negligible magnitude. We identify a nontrivial crossing between the strong and weak magnetic field regimes that reflects the competition between spin alignment and the energy required to sustain orbital motion. This nontrivial crossing temperature separates a spin-dominated regime from an inertia-dominated regime of magnetic field-induced rotation. These findings establish the EdH effect as a manifestation of angular momentum conservation in magnetized QCD matter.
\end{abstract}

\maketitle

\section{INTRODUCTION}

Ultra-relativistic heavy-ion collisions at the Relativistic Heavy-Ion
Collider (RHIC) and the Large Hadron Collider (LHC) provide compelling
evidence for the formation of a deconfined state of strongly interacting
matter known as the quark-gluon plasma (QGP), offering a unique
experimental window into Quantum Chromodynamics (QCD) under extreme
conditions of temperature ($T$), energy density ($\varepsilon$), and
baryon chemical potential ($\mu_{\rm B}$).
Over the past two decades, substantial progress has been made in
characterizing the bulk properties of the QGP, encompassing its equation
of state, transport coefficients, and collective flow
behavior~\cite{Kashlinsky:2018mnu,Heinz:2013th}.
Beyond these achievements, two phenomena intrinsic to non-central
heavy-ion collisions have opened fundamentally new avenues for probing
the QCD phase structure through the interplay of electromagnetic fields,
spin degrees of freedom, and collective dynamics.
These are the transient generation of the strongest magnetic fields known
in nature~\cite{Kharzeev:2007jp,Skokov:2009qp} and the experimental
observation of global hyperon
polarization~\cite{Becattini:2020ngo,STAR:2017ckg,ALICE:2019onw}, the
latter providing direct evidence for the substantial orbital angular
momentum deposited in the collision and its subsequent transfer to
the QGP medium via spin-orbit coupling.\\

In non-central heavy-ion collisions, the spectator protons generate intense yet transient magnetic fields, reaching magnitudes of order $eB \sim m_{\pi}^{2}$ at RHIC energies and substantially larger values at the LHC~\cite{Skokov:2009qp,Deng:2012pc}. These fields are extremely short-lived, decaying over proper times of approximately $5$--$10$~fm/$c$, governed by the finite electrical conductivity of the QGP and the rapidly expanding fireball geometry~\cite{Tuchin:2013apa}. Despite their transient nature, such magnetic fields introduce a new dynamical energy scale that can significantly modify the thermodynamic properties, transport coefficients, and collective dynamics of the QGP~\cite{Kharzeev:2013jha,Fukushima:2008xe}. A comprehensive understanding of the interplay between the magnetic field strength, temperature, and baryon chemical potential is therefore essential for a complete and consistent description of QGP evolution. Considerable theoretical and phenomenological effort has been devoted to characterizing QCD matter under external magnetic fields, leading to rigorous determinations of the thermodynamic and transport properties of both the hadronic phase and the deconfined medium under such conditions~\cite{Kharzeev:2012ph,Bali:2011qj,Bali:2012zg, Endrodi:2015oba,Pradhan:2021vtp,Sahoo:2023vkw,Goswami:2023eol}.\\

Simultaneously, the observation of global $\Lambda$ hyperon polarization
by the STAR Collaboration at RHIC~\cite{STAR:2017ckg} has established
that the QCD matter produced in heavy-ion collisions possesses the
largest fluid vorticity ever measured, with a magnitude of order
$\omega \sim 10^{21}$~s$^{-1}$.
This landmark observation has stimulated intense theoretical studies,
forging connections between orbital angular momentum, fluid vorticity,
and spin dynamics in QCD matter.
The prevailing theoretical paradigm holds that the large orbital angular
momentum carried by the colliding nuclei is partially converted into
fluid vorticity, which subsequently polarizes hadron spins through
spin-vorticity coupling~\cite{Becattini:2013fla,Becattini:2013vja}.
Within this framework, spin polarization is treated as a passive
response to the underlying fluid flow, evaluated either via the thermal
vorticity tensor at kinetic freeze-out or through covariant spin
transport equations~\cite{Huang:2020dtn,Becattini:2016gvu,%
Becattini:2022zvf}.
Despite substantial progress, several open issues persist, including
sensitivity to the choice of pseudogauge~\cite{Becattini:2012pp,%
Florkowski:2018fap}, the unresolved sign problem in longitudinal
polarization measurements~\cite{STAR:2019erd}, and the absence of
explicit total angular momentum conservation in most theoretical
implementations~\cite{Montenegro:2017rbu}.
Complementing these developments, considerable effort has recently been
directed toward understanding QCD matter under extreme rotation, with
particular focus on its effects on thermodynamic and transport
properties~\cite{Pradhan:2023rvf,Sahoo:2025fif,Dwibedi:2025boz,%
Dwibedi:2024amt,Padhan:2024edf,Braguta:2023kwl}.\\

In condensed-matter and atomic systems, the Einstein–de Haas (EdH) effect~\cite{EinsteinDeHaas1915} provides a direct demonstration of magnetomechanical coupling. In this effect, a change in the magnetization of a material induces a compensating mechanical rotation, ensuring conservation of total angular momentum. The Barnett effect~\cite{Barnett:1915uqc}, in contrast, is the reciprocal phenomenon, in which a mechanically rotating body develops a finite magnetization aligned with the axis of rotation. These phenomena have been extensively studied in ferromagnetic materials and, more recently, observed in quantum systems such as single-molecule magnets coupled to nanomechanical resonators~\cite{{Ganzhorn2016}} and Bose-Einstein condensates of europium atoms~\cite{Matsui}. In the context of heavy-ion collisions, the Barnett effect has been recently explored in QCD matter~\cite{Sahu:2025tmb}, where a rotating hadron gas is shown to generate a finite magnetic field comparable to the external fields present at RHIC and the LHC energies. Similarly, the EdH effect has also been studied with a static equilibrated hadron gas and found to generate finite rotation~\cite{Sahu:2026lbv}.
However, the EdH response has not yet been quantitatively investigated in the dynamically expanding QGP phase, which has the potential to significantly modify the vorticity inferred from final-state polarization measurements.\\

The hadronic phase of the fireball provides a suitable environment for the EdH effect as it contains a rich spectrum of particles with large magnetic moments, and it persists for approximately \(5\)--\(10\) fm/\(c\)~\cite{Sahu:2019tch}, experiencing residual magnetic fields near freeze-out. Moreover, the temperature region near the QCD crossover exhibits a pronounced variation in the magnetic susceptibility of the hadronic medium, reflecting the rapid change in degrees of freedom. A recent study~\cite{Sahu:2026lbv} demonstrated that even remnant magnetic fields at freeze-out can induce collective rotation of the fireball with angular velocities \(\omega_{\mathrm{EdH}} \sim 0.01\)--\(0.02\) GeV, comparable to typical estimates of fluid vorticity inferred from hyperon polarization. Crucially, this rotation arises solely from the magnetic field, without any initial vorticity input. The EdH effect in the HRG thus establishes hot QCD matter as a self-vortical magnetofluid in the hadronic phase, where spin and collective rotation are dynamically coupled. On the other hand, the QGP presents a fundamentally different environment from the hadron gas, in which quarks are deconfined, carry fractional electric charges, and have effective masses governed by the running coupling and thermal interactions rather than constituent masses. In the presence of strong magnetic fields, charged quarks exhibit Landau quantization, leading to a discrete spectrum of transverse-momentum states, and their spins experience Zeeman splitting due to the coupling of their magnetic moments to the external field. Unlike hadrons, whose magnetic moments are determined by their internal structure, quarks have elementary magnetic moments \(\mu_q = q/(2m_q)\) that depend inversely on their effective masses, which themselves vary with temperature and field strength~\cite{Xu:2020yag}. Furthermore, the QGP undergoes approximately boost-invariant longitudinal expansion, while the magnetic field decays exponentially with a decay constant determined by the QGP conductivity. Capturing these dynamical features is essential for a realistic estimate of the EdH effect across the entire fireball lifetime.\\

To address these challenges, we employ the quasiparticle model (QPM), which has been shown to successfully describe the thermodynamic properties of the QGP~\cite{Chandra:2011en,Peshier:1995ty}. In this framework, interacting quarks and gluons are treated as massive quasiparticles with thermal masses that incorporate the non-perturbative nature of QCD near the confinement transition. The model reproduces lattice QCD results for the equation of state at zero baryon density with high accuracy and has been systematically extended to finite chemical potential, finite magnetic fields, and out-of-equilibrium conditions~\cite{Levai:1997yx,Bluhm:2004xn}. However, a systematic study of spin density and the EdH effect in a dynamically evolving, magnetized QGP using the QPM has not yet been performed.\\

This paper is organized as follows. Section~\ref{sec:formalism} presents the theoretical framework, including the quasiparticle model for magnetized QGP, the treatment of Landau quantization and Zeeman splitting and the estimation of the EdH rotation. Section~\ref{sec:results} presents our results, discusses the physical implications of our findings, addresses the role of dynamical spin relaxation, and provides experimental predictions for upcoming measurements at RHIC, LHC, and future facilities. Section~\ref{sec:summary} summarizes our conclusions and outlines directions for future work.

\section{Formalism}
\label{sec:formalism}

In this section, we present the formalism for estimating the EdH effect in a magnetized, expanding QGP using a quasiparticle model. We assume a simplistic boost-invariant longitudinal Bjorken expansion of the plasma and corresponding temperature cooling is given as~\cite{Bjorken:1982qr},

\begin{equation}
T(\tau) = T_0 \left(\frac{\tau_0}{\tau}\right)^{1/3},
\end{equation}
where, $\tau$ is the proper time, $T_0 = 350$ MeV, is the initial temperature at thermalization time $\tau_0 = 0.5$ fm. The deconfinement critical temperature is taken as $T_c = 0.155~\text{GeV}$~\cite{HotQCD:2014kol}.

The external magnetic field generated by spectator protons decays due to the expansion of the fireball~\cite{Skokov:2009qp}. This decay is modeled through a phenomenological exponential ansatz~\cite{Voronyuk:2011jd,Deng:2012pc},
\begin{equation}
eB(\tau) = eB_{0} \exp\left(-\frac{\tau}{\tau_{\rm B}}\right),
\label{eq:eB_decay}
\end{equation}
where $eB_{0}$ denotes the initial field strength at $\tau = 0$ and $\tau_{\rm B}$ is the 
characteristic decay constant. Two representative initial field strengths are considered,
$eB_{0} = m_{\pi}^{2}$ and $eB_{0} = 3m_{\pi}^{2}$, with the pion mass taken as $m_{\pi} = 0.140$~GeV.
To account for uncertainties arising from the finite electrical conductivity of the QGP, the decay 
constant is varied in the range $\tau_{\rm B} = 5 \pm 2$~fm~\cite{Tuchin:2013apa}.\\

The QCD running coupling constant in the presence of the magnetic field is defined as~\cite{Ayala:2018wux,Singh:2025geq};

\begin{equation}
\alpha_s(T,\mu_{\rm B},eB) = \frac{\alpha_s^{(0)}}{1 + \gamma \alpha_s^{(0)} \ln\left(\frac{T^2 + \mu_{\rm B}^2/g^2}{T^2 + \mu_{\rm B}^2/g^2 + |eB|}\right)},
\end{equation}

where

\[
\gamma = \frac{11N_c - 2N_f}{12\pi}, \qquad g^2 = 4\pi\alpha_s^{(0)},
\]

with $N_c = 3$ colors and $N_f = 3$ flavors ($u$, $d$, $s$). The zero-field coupling is approximated by a constant, $\alpha_s^{(0)} = 0.5$, which reproduces typical values in the temperature range of interest. The QCD scale $\Lambda$ has been absorbed into the normalization, and the logarithm encodes the magnetic field suppression of the coupling at large $eB$.\\

 Further, medium interaction is characterized using QPM, which treats interacting quarks and gluons as massive quasiparticles with thermal masses~\cite{Gorenstein:1995vm,Peshier:1999ww}. For quarks, the thermal mass in the presence of a magnetic field is expressed as,

\begin{equation}
m_{q,T}^2 = \frac{N_c^2-1}{8N_c} g^2 \left(T^2 + \frac{\mu_{\rm B}^2}{9\pi^2}\right).
\end{equation}

The effective quark mass is then constructed to interpolate between the bare mass $m_0$ and the thermal mass,

\begin{equation}
m_q^2 = m_0^2 + \sqrt{2} m_0 m_{q,T} + m_{q,T}^2.
\end{equation}

This form ensures that $m_q \to m_0$ as $T \to 0$ and $m_q \to m_{q,T}$ at high temperature. For the bare masses we use $m_{u0} = 0.0023~\text{GeV}, \quad m_{d0} = 0.0048~\text{GeV}, \quad m_{s0} = 0.095~\text{GeV}$. For gluons, the effective mass is;

\begin{equation}
m_g^2 = \frac{N_c}{6} g^2 T^2 \left[1 + \frac{1}{6}\left(N_f + \frac{\mu_{\rm B}^2}{g^2 T^2}\right)\right].
\end{equation}

In the presence of a magnetic field, charged quarks occupy discrete Landau levels. The transverse energy spectrum is quantized as,

\[
E_\perp^2 = m_q^2 + 2|q_f eB| n, \qquad n = 0,1,2,\ldots,
\]

where $q_f$ is the fractional quark charge ($q_u = 2/3$, $q_d = -1/3$, $q_s = -1/3$). The Landau level degeneracy is $\alpha_n = 2 - \delta_{n0}$.

Intrinsic spin effects are incorporated via Zeeman splitting,

\[
\Delta E_{\text{Zeeman}} = -s \mu_f B, \qquad s = \pm 1,
\]

where the quark magnetic moment is

\[
\mu_f = \frac{q_f}{2m_q}.
\]

Following these, the quasiparticle dispersion relation for a quark in a magnetic field is modified as~\cite{Sahu:2026lbv},

\[
E_{n,s}(k_z) = \sqrt{k_z^2 + m_q^2 + 2|q_f eB| n} - s \mu_f B.
\]

Gluons, being neutral particles, remain unaffected by the magnetic field at leading order and retain the continuum dispersion $E_g(\mathbf{k}) = \sqrt{|\mathbf{k}|^2 + m_g^2}$.\\

The dispersion used here separates orbital Landau quantization and explicit Zeeman splitting. Although the Dirac equation with $g=2$ naturally yields a combined spin-orbit term, our explicit separation is a deliberate choice: it allows us to isolate the spin-only contribution to the pressure derivative, which is necessary for computing the EdH rotation. This approximation is valid within our quasiparticle model, where $m_q$ does not depend on spin orientation.\\

The distribution functions for quarks (antiquarks) and gluons are given
by the Fermi-Dirac and Bose-Einstein distributions, respectively,
\begin{equation}
f_{\rm FD}(E) = \frac{1}{e^{(E-\mu)/T}+1}, \qquad
f_{\rm BE}(E) = \frac{1}{e^{E/T}-1},
\label{eq:dist_func}
\end{equation}
where the effective quark chemical potential is $\mu = \mu_{\rm B}/3$,
with $\mu_{\rm B}$ being the baryon chemical potential.
The quark contribution to the energy density in the presence of an
external magnetic field, incorporating Landau quantization, reads
\begin{equation}
\varepsilon_{f} = N_{c} \sum_{n=0}^{\infty} \alpha_{n}
\sum_{s=\pm 1} \frac{|q_{f}\,eB|}{2\pi^{2}}
\int_{-\infty}^{\infty} dk_{z}\,
E_{n,s}(k_{z})\, f_{\rm FD}\!\left(E_{n,s}\right),
\label{eq:eps_quark}
\end{equation}
where $N_{c} = 3$ is the number of colors, $\alpha_{n} = 2 - \delta_{n0}$
is the Landau level degeneracy factor, $q_{f}$ is the electric charge of
quark flavor $f$, and $E_{n,s}(k_{z})$ is the quark dispersion relation
in the $n$-th Landau level with spin projection $s$.
The gluon contribution is evaluated in the continuum limit,
\begin{equation}
\varepsilon_{g} = g_{d} \int \frac{d^{3}k}{(2\pi)^{3}}\,
E_{g}(\mathbf{k})\, f_{\rm BE}\!\left(E_{g}\right),
\label{eq:eps_gluon}
\end{equation}
where the gluon degeneracy factor $g_{d} = 16$ accounts for $8$ color
states and $2$ helicity states. The net energy density of the QGP is then obtained by summing over  quark flavors $f \in \{u, d, s\}$, their antiquark
counterparts, the gluon sector, and the perturbative vacuum contribution,
\begin{equation}
\varepsilon = \sum_{f=u,d,s}
\left(\varepsilon_{f} + \varepsilon_{\bar{f}}\right)
+ \varepsilon_{g} + B_{\rm vac},
\label{eq:eps_total}
\end{equation}
here $B_{\rm vac} = 0.0028$~GeV$^{4}$ is the bag constant encoding
the non-perturbative vacuum pressure.\\

Similarly, the pressure of the system due to quark (antiquark) is defined by taking the sum over quark flavors, along with the longitudinal pressure for fermions in Landau levels, expressed as

\begin{equation}
P_f = N_c \sum_{n,s} \alpha_n \frac{|q_f eB|}{2\pi^2} \int dk_z \, \frac{k_z^2}{\omega_{n,s}} f_{\text{FD}}(\omega_{n,s})
\end{equation}

The corresponding gluonic contribution to the medium pressure is:
\begin{equation}
P_g
=
g_d
\int \frac{d^3k}{(2\pi)^3}
\frac{k^2}{3\omega_g}
f_{\rm BE}(\omega_g) 
\end{equation}

where, $\omega_g = \sqrt{k^2 + m_g^2},$ accounts for the dispersion relation. Since the gluon dispersion relation contains no explicit dependence on the magnetic field $B$, the gluon pressure does not acquire a direct magnetic-field dependence. The net thermodynamic pressure of the medium can be written as;

\begin{equation}
P = \sum_{f=u,d,s}\left(P_{f} + P_{\bar{f}}\right)
+ P_{g} - B_{\rm vac},
\label{eq:pressure_total}
\end{equation}

where the bag constant $B_{\rm vac}$ encodes the non-perturbative
vacuum contribution, consistent with Eq.~\eqref{eq:eps_total}.\\

The bulk magnetization of the medium is defined through the standard
thermodynamic relation,
\begin{equation}
M_{z} = \left(\frac{\partial P}{\partial B}\right)_{T,\mu}.
\label{eq:magnetization}
\end{equation}

In the context of the Einstein--de~Haas (EdH) effect, however, it is
essential to distinguish between the two physically distinct
contributions to $M_{z}$. The orbital contribution, associated with
the cyclotron motion of charged particles across Landau levels,
corresponds to internal circulating currents that do not exert a net
macroscopic torque on the system. In contrast, the intrinsic spin
contribution arising from the Zeeman coupling of quark spins to the
external field, carries angular momentum and can, upon
field quench, be transferred to the medium as global mechanical
rotation via the EdH mechanism.

To isolate the spin-only contribution, we compute the pressure derivative with respect to the magnetic field, retaining only terms that explicitly depend on the Zeeman splitting. This is achieved by keeping only $d(\Delta E_{\text{Zeeman}})/dB$. The resulting spin-only pressure derivative is denoted $(\partial P/\partial B)_{\text{spin}}$.

The spin density is then related to this derivative via the thermodynamic relation connecting magnetic moment and spin,

\[
S_z = \frac{2m}{ge} \left(\frac{\partial P}{\partial B}\right)_{\text{spin}},
\]

The Einstein--de Haas rotation originates from the
conservation of total angular momentum,
\begin{equation}
J_{\rm tot}=J_{\rm spin}+J_{\rm orb}.
\end{equation}
In the presence of a magnetic field, spin alignment
generates a net intrinsic angular momentum density
$S_z$. Conservation of total angular momentum then
requires the development of a compensating orbital
rotation of the system.

For a longitudinally expanding fireball with approximate cylindrical symmetry, the total spin angular momentum contained within a volume $V$ is
\begin{equation}
J_{\rm spin}=S_z V,
\end{equation}
while the orbital contribution is approximated by
\begin{equation}
J_{\rm orb}=I\omega_{\rm EdH},
\end{equation}
where $\omega_{\rm EdH}$ denotes the effective global
angular velocity induced by the EdH mechanism. The
moment of inertia is given as,
\begin{equation}
I=\kappa \varepsilon V R^2,
\end{equation}
with $\varepsilon$ the total energy density, $R$ the transverse fireball radius, and $\kappa$ a geometric factor encoding the radial energy-density profile. In relativistic systems,
the rotational inertia is governed by the energy density rather than the rest-mass density, motivating the present effective form.

Assuming that the EdH-induced rotation remains small and treating the expansion as slow compared to the spin-alignment timescale, the angular momentum conservation implies
\begin{equation}
\partial_\tau
\left(
S_z V + I\omega_{\rm EdH}
\right)
\simeq 0.
\end{equation}
Neglecting higher-order corrections from the explicit
time dependence of $I(\tau)$ then yields
\begin{equation}
\omega_{\rm EdH}
\simeq
-\frac{S_zV}{I}
=
-\frac{S_z}{\kappa \varepsilon R^2}.
\end{equation}

Gluons are electrically neutral and couple to an external
electromagnetic field only indirectly, through quark-loop corrections
to the gluon polarization tensor and the effective quark--gluon
interaction~\cite{Ayala:2019akk}. As a result, a uniform background
magnetic field induces Zeeman splitting exclusively in the quark sector,
and the gluon spin contribution to the total spin density $S_{z}$
remains parametrically suppressed relative to the quark contribution
at leading order. The present analysis, therefore, focuses on the quark
spin sector as the dominant source of spin-to-orbital angular momentum
conversion via the EdH mechanism.\\

Within this framework, the induced angular velocity carries a negative sign consistent with the fundamental EdH relation $\Delta J_{\rm rotation} = -\Delta J_{\rm spin}$, reflecting the
transfer of spin angular momentum released upon field quench into global rotation of the fireball.
For the nominal calculation, the fireball radius and geometric
inertia factor are taken as $R = 5$~fm and $\kappa = 0.5$,
respectively, giving a moment of inertia
$\mathcal{I} = \kappa M R^{2}$. The sensitivity of the results to the considered fireball geometry is
assessed by varying $\kappa \in [0.3,\, 1.0]$.
The EdH angular velocity $\Omega_{\rm EdH}(\tau)$ carries an implicit
proper-time dependence through the evolving temperature profile
$T(\tau)$ and the decaying magnetic field $B(\tau)$, dynamically
encoding the interplay between fireball cooling, field quench, and
spin-to-rotation conversion. The rigid-rotation approximation serves as an effective description of the global rotational response of the expanding fireball, valid on length scales large compared to the mean free path of the medium constituents.

\section{Results and Discussions}
\label{sec:results}

\begin{figure*}[htbp]
    \centering
    \includegraphics[width=0.95\textwidth]{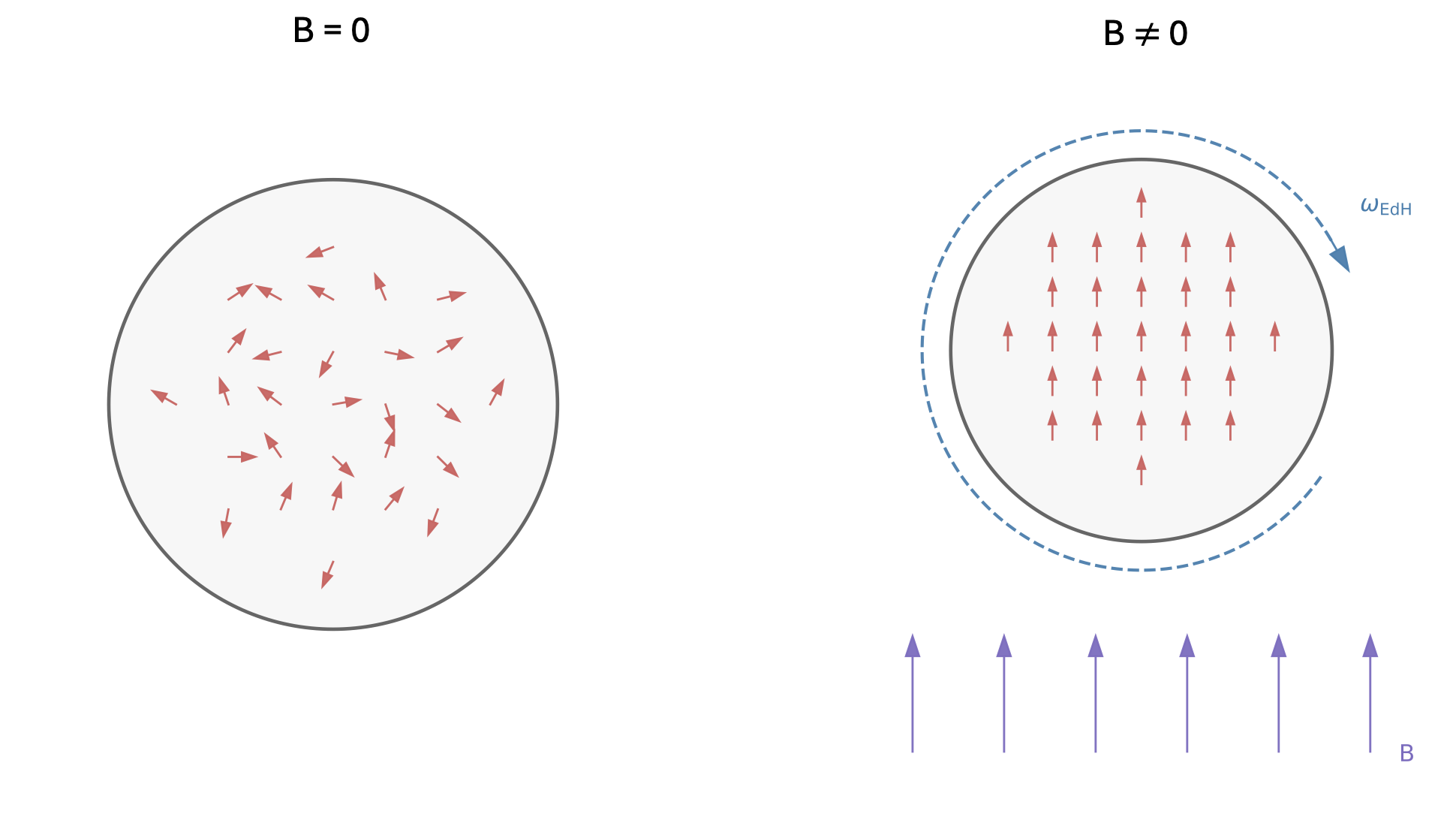}
    \caption{A schematic representation of how magnetic field creates a spin alignment, which in turn creates a rotation to conserve
the total angular momentum. The figure is taken from Ref.~\cite{Sahu:2026lbv}.}
    \label{fig1}
\end{figure*}


The schematic diagram shown in Fig.~\ref{fig1} illustrates the Einstein--de Haas mechanism for quarks (antiquarks) in the medium, which initially do not possess global rotation. In the absence of an external magnetic field, quark spins are randomly oriented, resulting in a vanishing net spin angular momentum, as depicted in the left panel. When a magnetic field is
applied, the quark magnetic moments tend to align with the field, producing a nonzero spin polarization and hence a finite spin angular momentum. Conservation of total angular momentum then requires the medium to acquire a compensating orbital angular momentum, leading to a collective rotation with angular velocity $\omega_{\rm EdH}$ opposite to the direction of the induced spin
alignment. The magnitude of this rotational response is determined by the magnetization and spin polarization of the quarks and antiquarks present in the medium.

\begin{figure}[htbp]
    \centering
    \includegraphics[width=0.45\textwidth]{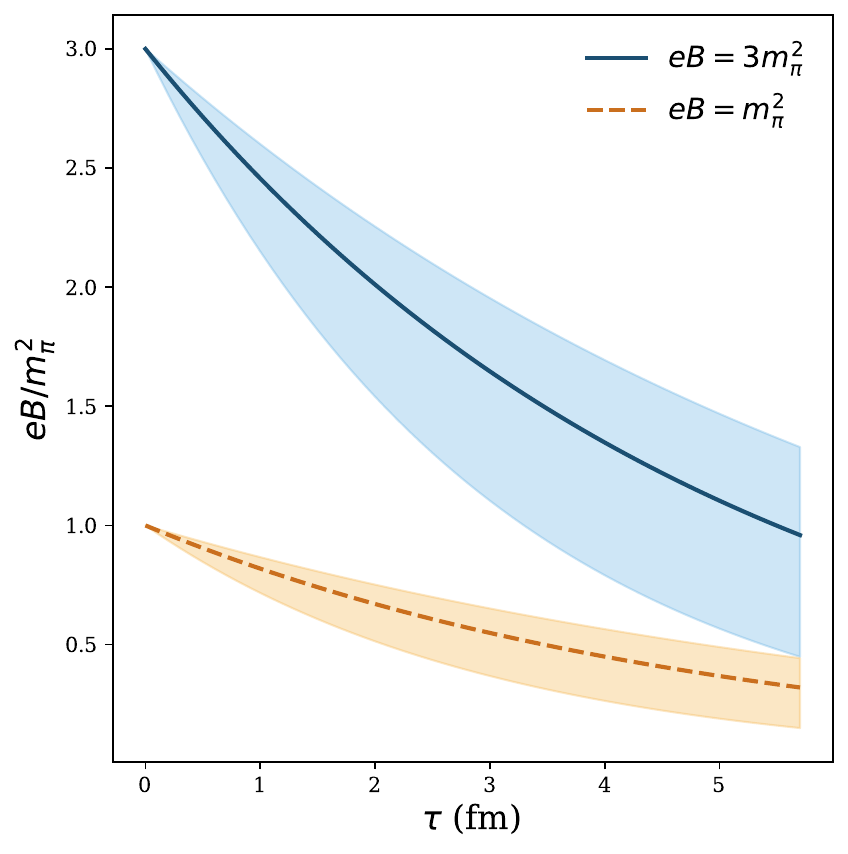}
    \caption{Magnetic field evolution with an exponential decay profile for two initial values of $eB$. The shaded band is for uncertainties in the magnetic field decay timescale ($\tau_{B}$).}
    \label{fig2}
\end{figure}

The magnetic field decays exponentially with proper time ($tau$) as shown in fig.~\ref{fig2}. Two initial magnetic field values are used, corresponding to strong ($eB = 3m_\pi^2$) and weak ($eB = m_\pi^2$) magnetic field. The $\tau_{\rm B}$ uncertainty of $3$--$7$ fm leads to increasing relative errors, reaching $\sim 50\%$ at $\tau = 5$ fm for the fastest-decay scenario. Notably, the magnetic field remains significant throughout the QGP lifetime, ensuring that magnetic spin polarization persists through the phase transition and into the hadronic phase. The cooling timescale coincides with the magnetic field decay constant $\tau_{\rm B} = 5$ fm, creating a favorable regime near $\tau \approx 4$--5 fm where the temperature is sufficiently low for efficient spin alignment ($T \sim T_c$) while the magnetic field remains strong enough to produce significant spin polarization. This temporal coincidence maximizes the Einstein-de Haas effect in realistic heavy-ion collision scenarios. While more sophisticated magnetohydrodynamic descriptions exist, the exponential form captures the leading effect of finite conductivity and allows a transparent estimate of the EdH response. With the evolving $T$ and $eB$, we proceed towards further calculations.\\

\begin{figure*}[htbp]
    \centering
    \includegraphics[width=0.45\textwidth]{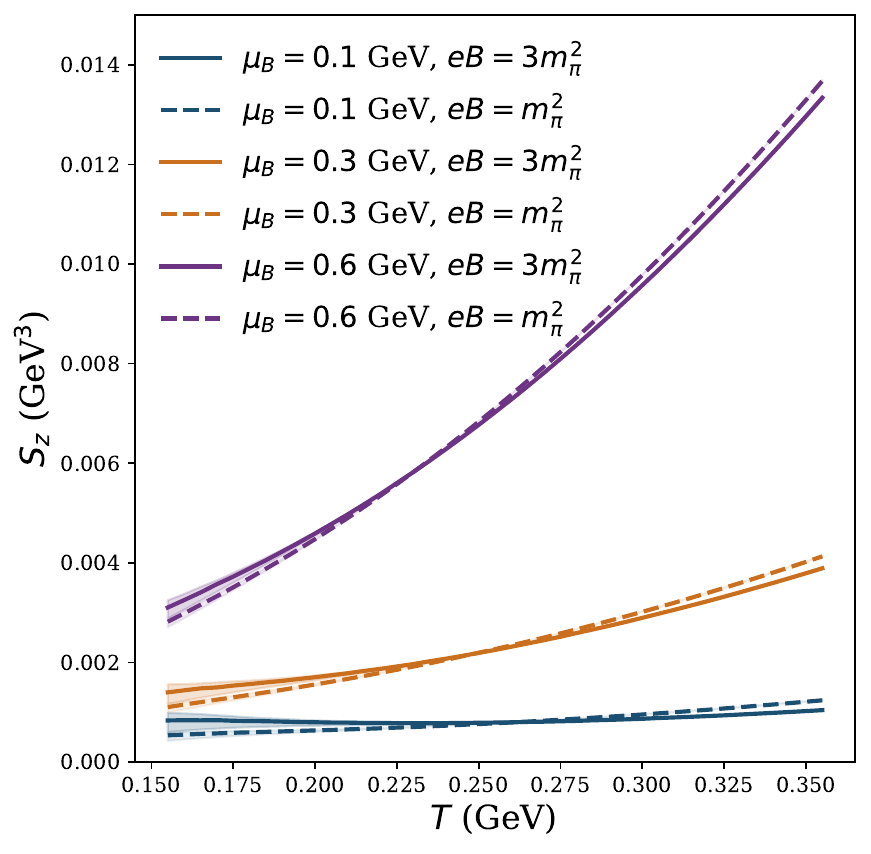}
    \includegraphics[width=0.45\textwidth]{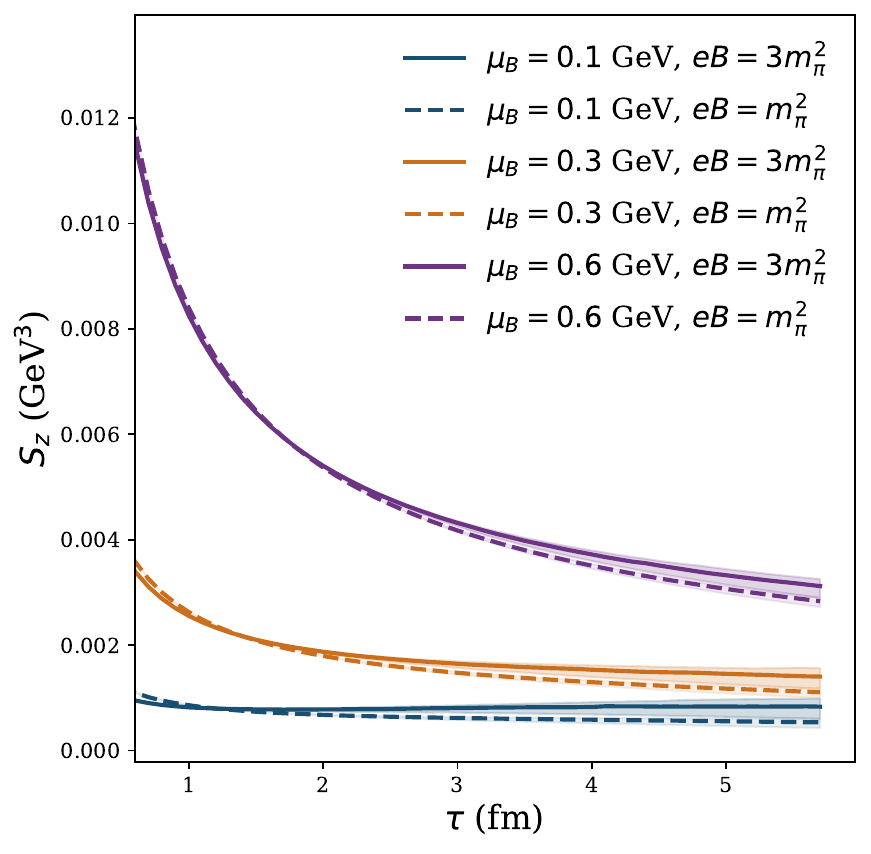}
    \caption{Spin density of the system as a function of temperature (left panel) and proper time ($\tau$) for different baryon chemical potential and initial magnetic field values. }
    \label{fig3}
\end{figure*}

Figure~\ref{fig3} shows the spin density $S_z$ of the system as a function of temperature (left panel) and proper time (right panel). This is estimated for both strong and weak initial magnetic field cases, with three distinct values of the baryon chemical potential, to investigate the density dependence and to get an idea of the QCD phase space. 
The $S_z$ exhibits a strong dependence on temperature and proper time, governed by the interplay between quark effective masses, thermal fluctuations, Pauli blocking, and magnetic field decay. For $\mu_{\rm B} = 0.1$ GeV, $S_z$ remains almost flat within uncertainties. In contrast, for $\mu_{\rm B} = 0.3$ and 0.6 GeV, $S_z$ decreases with cooling, as Pauli blocking suppresses additional spin polarization in the dense Fermi sea. The magnetic field strength affects $S_z$ by up to 50\% at low $\mu_{\rm B}$, but this enhancement saturates at high $\mu_{\rm B}$ where the system is already highly polarized. The medium evolution reflects the competing effects of Bjorken cooling and exponential decay of the magnetic field, which universally suppresses $S_z$ at later times. This non-trivial behavior demonstrates that the EdH effect is most pronounced in baryon-rich, near-critical matter with moderate magnetic fields.\\

\begin{figure*}[htbp]
    \centering
    \includegraphics[width=0.45\textwidth]{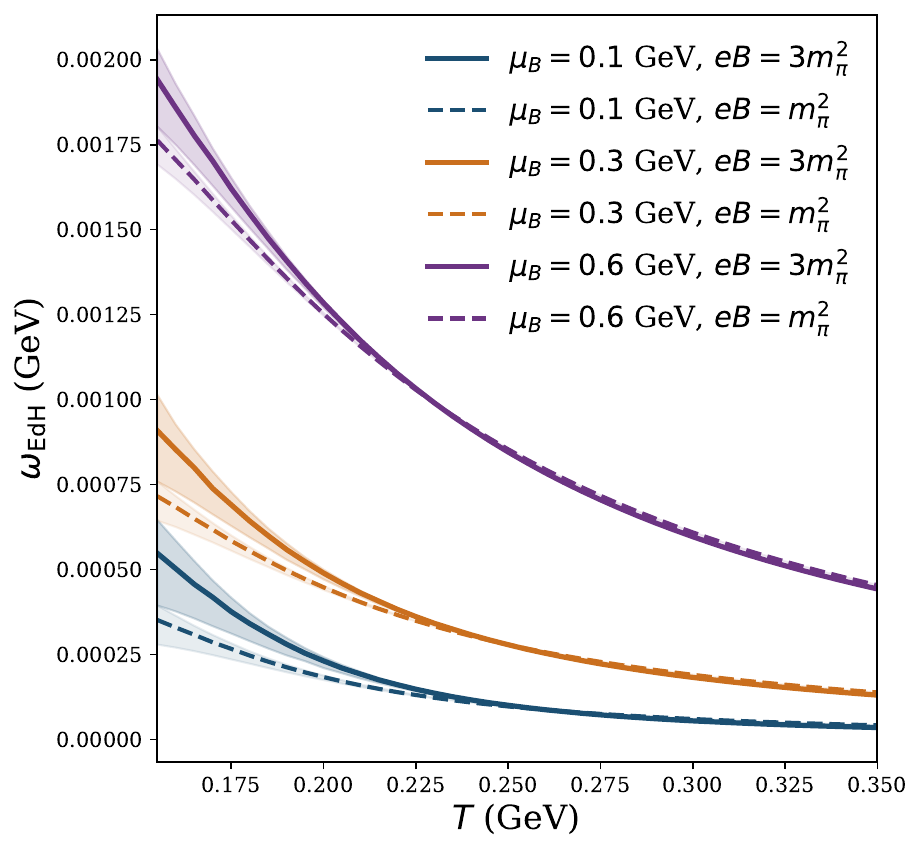}
    \includegraphics[width=0.42\textwidth]{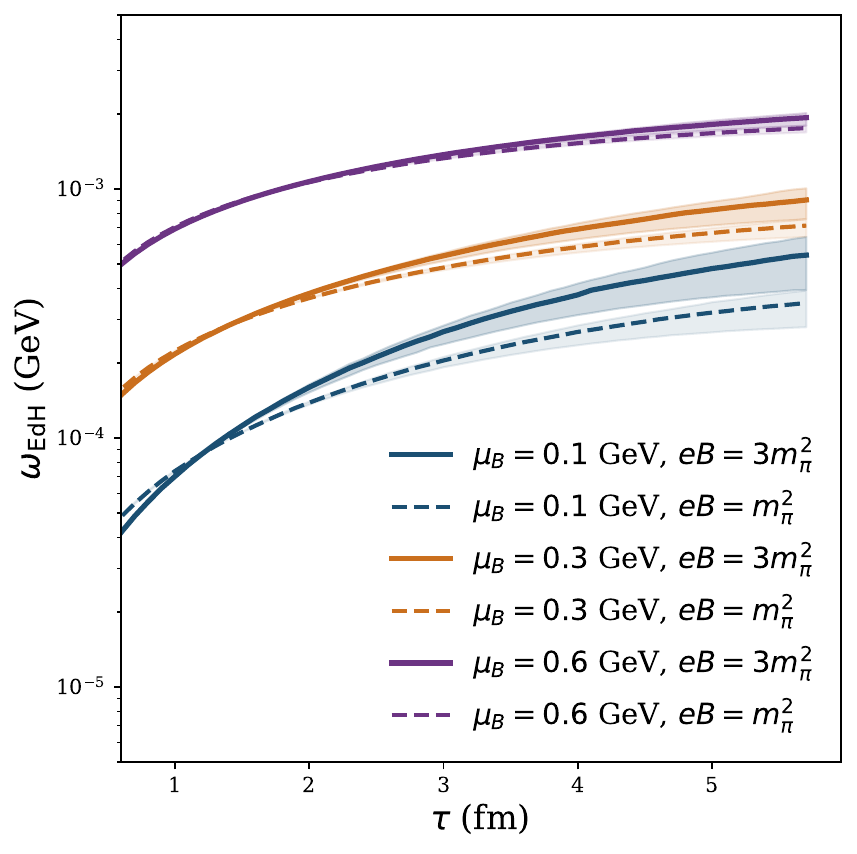}
    \caption{Induced rotation ($\omega_{EdH}$) as a function of temperature (left panel) and proper time ($\tau$) for different baryon chemical potential and initial magnetic field values.}
    \label{fig4}
\end{figure*}

Figure~\ref{fig4} shows the absolute value of induced rotation $\omega_{\mathrm{EdH}}$ (we have removed the negative sign convention here) as a function of temperature (left panel) and proper time (right panel). Our estimations reveal two key features of the Einstein-de Haas effect in magnetized QGP. First, the angular velocity $\omega_{\mathrm{EdH}}$ exhibits a characteristic temperature dependence: it peaks near the critical temperature $T_c \approx 0.155$ GeV and decreases monotonically as temperature increases. This behavior arises from the competition between spin polarization and thermal fluctuations near $T_c$, where quarks have large effective masses that enhance the Zeeman splitting, leading to efficient spin alignment. On the other hand, at higher temperatures, thermal fluctuations reduce the net spin density $S_z$. We observe a crossing between strong and weak magnetic field curves at $T \approx 0.28$ GeV for $\mu_{\rm B} = 0.1$ GeV. This crossing occurs at lower temperatures for higher $\mu_{\rm B}$ cases. Below this crossing, larger magnetic fields produce faster rotation ($\omega_{\mathrm{strong}} > \omega_{\mathrm{weak}}$) because the enhanced spin polarization dominates; above the crossing, the trend reverses ($\omega_{\mathrm{strong}} < \omega_{\mathrm{weak}}$) as relativistic effects suppress the spin response while the orbital energy density from Landau quantization becomes comparable. Second, the proper time evolution follows the Bjorken expansion with $\omega_{\mathrm{EdH}}$ increasing rapidly with $\tau$. At early times ($\tau \sim \tau_0 = 0.5$ fm), the system is hot and highly magnetized, producing negligible rotation due to thermal spin depolarization. As the system expands and cools, $\omega_{\mathrm{EdH}}$ rises, reaching a maximum around $T_c$. This yields a non-monotonic $\tau$-dependence that reflects the interplay between temperature-driven spin polarization and field decay. The resulting angular velocities are in the range $0.05$--$2$ MeV, within the sensitivity of polarization measurements at the RHIC and LHC energies.\\

\begin{figure}[htbp]
    \centering
    \includegraphics[width=0.45\textwidth]{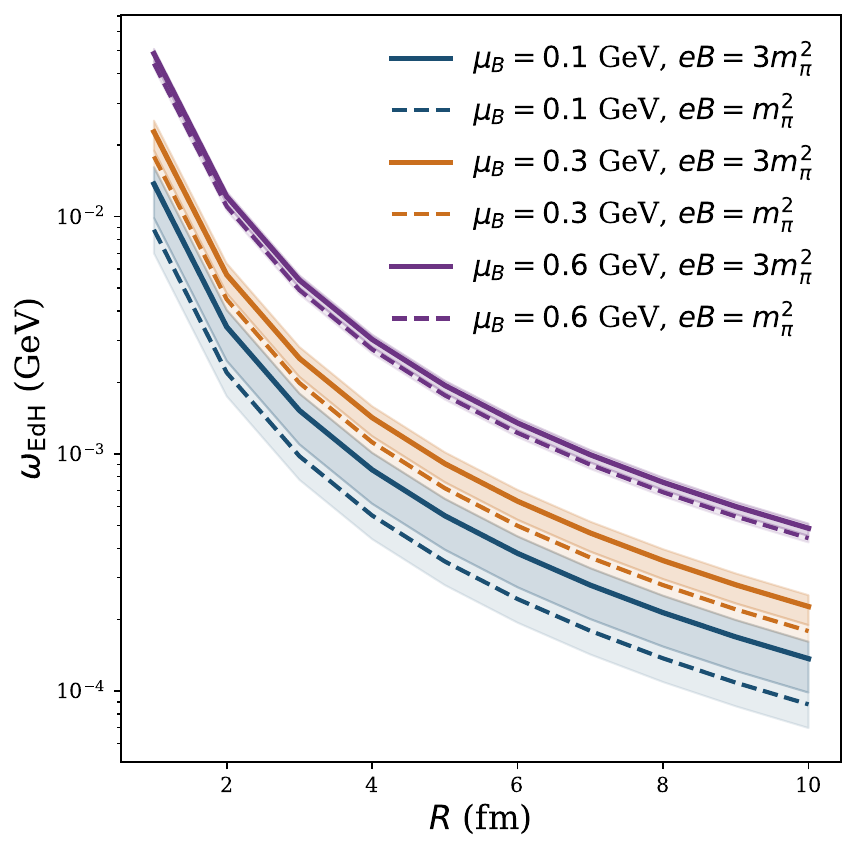}
    \caption{Induced rotation ($\omega_{EdH}$) as a function of system radius ($R$) for different baryon chemical potential and initial magnetic field values at the transition temperature ($T_{c}$ = 155 MeV).}
    \label{fig5}
\end{figure}

A particularly nontrivial feature of our results is the emergence of a crossing temperature $T_{\rm cross}$ separating two distinct dynamical regimes of the Einstein--de Haas response. This inversion originates from the competition between spin polarization and the rapidly increasing orbital energy associated with Landau quantization in the relativistic plasma. At high temperatures, thermal motion suppresses the net spin alignment while the magnetic field simultaneously increases the energy density and effective rotational inertia of the system. Consequently, the enhancement of the denominator in $\omega_{\rm EdH}\propto S_z/\varepsilon$ outpaces the growth of the spin density itself, driving the system into an inertia-dominated regime where magnetic fields become less efficient at producing collective rotation. The crossing, therefore, marks a dynamical spin-to-inertia crossover in magnetized QCD matter, demonstrating that stronger magnetic fields do not necessarily imply stronger vorticity generation in the quark--gluon plasma. This behavior has important phenomenological implications, suggesting that the Einstein--de Haas effect may be comparatively more significant at intermediate temperatures and baryon-rich collision energies than in the hottest plasma produced at the top of the LHC energies.\\

Figure~\ref{fig5} shows the absolute value of induced angular velocity $\omega_{\mathrm{EdH}}$ as a function of system radius. $\omega_{\mathrm{EdH}}$ exhibits a clean $\propto 1/R^2$ scaling with the fireball radius, arising from angular momentum conservation. For a large system size, with $R = 10$ fm, the induced rotation is smaller, owing to the fact that the spin density becomes smaller. However, a larger baryon chemical potential can still give a non-negligible induced rotation through the EdH effect. A clear magnetic field dependence is also observed, where throughout the R dependence, a strong magnetic field gives a higher $\omega_{\mathrm{EdH}}$ for the corresponding $\mu_{\rm B}$. Although the magnetic field dependence seems to be decreasing with an increase in $\mu_{\rm B}$. More peripheral collisions or smaller systems with small $R$ produce larger instantaneous rotation.\\

Our current estimations yield substantial $\omega_{\mathrm{EdH}}$ for typical RHIC conditions. The induced rotation increases strongly with baryon chemical potential, indicating that the EdH effect is most pronounced at lower collision energies (RHIC Beam Energy Scan, FAIR, NICA). The fireball radius scaling $\omega \propto R^{-2}$ offers a distinct experimental signature that can discriminate EdH rotation from conventional vorticity. The present QPM study complements our earlier HRG analysis by establishing that the EdH effect operates in both the partonic and hadronic phases of the fireball. Depending on the relative orientation between the
magnetic-field-induced spin alignment and the underlying hydrodynamic vorticity, the EdH contribution may either enhance or partially suppress the net observed rotation. This backreaction implies that the vorticity extracted from polarization measurements, $\omega_{\mathrm{total}} = \omega_{\mathrm{hydro}} + \omega_{\mathrm{EdH}}$, may be quantitatively modified relative to the initial hydrodynamic vorticity.\\

One important nuance here is that, in a fluid, angular momentum conservation does not enforce rigid rotation. The conservation law is local: $\partial_\mu \left( J^{\mu}_{\text{spin}} + J^{\mu}_{\text{orb}} \right) = 0$, which does not a priori require rigid-body rotation. In principle, the orbital angular momentum can be carried by differential flow gradients (shear, non-uniform vorticity, or radial expansion) rather than a single global angular velocity. Nevertheless, for the purpose of an order-of-magnitude estimate, we assume that spin-to-orbital conversion results in an effective global rotation characterized by a single angular velocity $\omega_{\mathrm{EdH}}$. This approximation is motivated by several considerations. Firstly, the fireball is approximately cylindrically symmetric. Secondly, the QGP has non-zero shear viscosity ($\eta/s \sim 0.1$--$0.2$), which tends to homogenize vorticity on a timescale $\tau_{\mathrm{vis}} \sim \eta/(\epsilon R^2) \sim 1$--$2$ fm/$c$, comparable to the expansion time; and finally, the uniform external magnetic field produces a uniform spin polarization, favoring a uniform rotational response in equilibrium. We therefore define $\omega_{\mathrm{EdH}}$ as the effective global angular velocity that would carry the same total orbital angular momentum if the fireball rotated rigidly with the measured energy density profile: $J_{\mathrm{orb}} \equiv \int d^3r \, (\mathbf{r} \times \mathbf{p}) = I \, \omega_{\mathrm{EdH}}$, with the moment of inertia of a rigid body with the same profile. Within this effective approach, $\omega_{\mathrm{EdH}}$ should be interpreted as a volume-averaged induced vorticity $\langle \omega \rangle$, which is the relevant quantity for comparison with experimental polarization measurements at freeze-out. The actual local vorticity field $\boldsymbol{\omega}(\mathbf{r})$ may be non-uniform, and its determination requires a full spin-hydrodynamic treatment, which is left for future work.\\

The present study also assumes that the spin polarization instantaneously follows its local thermodynamic equilibrium
value determined by the magnetic field and temperature. In reality, spin alignment in the QGP occurs over a finite spin relaxation timescale $\tau_s$, governed by spin-flip
scatterings and medium interactions. A more complete description would require solving a dynamical spin transport equation of the form\\
\begin{equation}
\frac{dS_z}{d\tau}
=
-\frac{S_z-S_{\rm eq}}{\tau_s},
\end{equation}
where $S_{\rm eq}$ is the equilibrium spin density obtained
from the quasiparticle model. In the fast relaxation limit
$\tau_s \ll \tau_{\rm exp}$, with $\tau_{\rm exp}$ the
hydrodynamic expansion timescale, the present equilibrium
treatment is recovered. Conversely, if $\tau_s$ becomes
comparable to the QGP lifetime, the actual spin density
may lag behind the equilibrium value, suppressing the
resulting Einstein--de Haas rotation. The quantitative
determination of $\tau_s$ in magnetized QCD matter,
and its coupling to spin hydrodynamic evolution,
remains an important direction for future investigation.

\section{Conclusion}
\label{sec:summary}

In this work, we have investigated the Einstein--de Haas effect in a magnetized quark--gluon plasma using a quasiparticle model under an external magnetic field, incorporating Bjorken expansion, and time-dependent
magnetic-field decay. The resulting EdH-induced angular velocities are found to be comparable in scale to typical vorticity values inferred from polarization measurements in low-energy relativistic heavy-ion collisions. The induced rotation
increases significantly with baryon chemical potential and exhibits a characteristic
$\omega_{\mathrm{EdH}} \propto R^{-2}$ dependence on the fireball radius arising from angular momentum conservation. A particularly nontrivial result is the emergence of a crossing temperature, where the field-strength dependence of the EdH response reverses. Below this temperature, stronger magnetic
fields enhance spin alignment and produce larger collective rotation, whereas at higher temperatures, the increasing orbital energy and effective rotational inertia
suppress the EdH response. This behavior signals a dynamical crossover between a spin-dominated regime and an inertia-dominated regime of magnetized QCD matter. The coincidence between the magnetic-field decay
timescale and the cooling time toward the transition temperature implies that substantial spin alignment may
survive until hadronization, allowing the rotational response generated in the QGP to influence the hadronic phase. The present results, therefore, support a unified
picture of spin-to-orbital angular momentum conversion across the full evolution of the fireball.\\

Although the present calculation is based on an equilibrium treatment with simplified longitudinal expansion, the predicted energy, system-size, and magnetic-field dependencies provide concrete phenomenological signatures that can be explored in future phenomenological studies and experimental measurements at RHIC, LHC, FAIR, and NICA. The present mechanism may also provide an additional channel for spin-orbital angular momentum exchange in relativistic spin hydrodynamics, suggesting that magnetic-field-driven spin conversion could represent a previously unexplored component of the rotational dynamics of strongly interacting matter.

\section*{Acknowledgement}
DS acknowledges the support from the postdoctoral fellowship of the DGAPA UNAM.
\\

\bibliographystyle{apsrev4-2}
{}

\end{document}